\documentclass[article,notoc]{JHEP3}

\usepackage{amsmath,amssymb}
\usepackage[dvips]{graphics}
\usepackage{epsfig}

\newcommand{\be}{\begin{equation}}
\newcommand{\ee}{\end{equation}}
\newcommand{\bea}{\begin{eqnarray}}
\newcommand{\eea}{\end{eqnarray}}
\newcommand{\bean}{\begin{eqnarray*}}
\newcommand{\eean}{\end{eqnarray*}}
\newcommand{\beq}{\begin{eqnarray}}
\newcommand{\eeq}{\end{eqnarray}}

\def\a{\alpha'}
\def\R{{\mathcal R}}

\title{On the temperature dependence of the absorption cross section for black holes in string theory}

\author{Filipe Moura
\\
Centro de Matem\'atica da Universidade do Minho, \\Escola de Ci\^encias, Campus de Gualtar, \\4710-057 Braga, Portugal\\
\\
\email{fmoura@math.uminho.pt}
}

\abstract{We study the low frequency absorption cross section of spherically symmetric nonextremal $d$-dimensional black holes. In the presence of $\a$ corrections, this quantity must have an explicit dependence on the Hawking temperature of the form $1/T_H$. This property of the low frequency absorption cross section is shared by the D1-D5 system from type IIB superstring theory already at the classical level, without $\a$ corrections. We apply our formula to the simplest example, the classical $d$-dimensional Reissner-Nordstr\"om solution, checking that the obtained formula for the cross section has a smooth extremal limit. We also apply it for a $d$-dimensional Tangherlini-like solution with $\a^3$ corrections.}

\begin{document}



\vfill

\eject

\section{Introduction and summary}
\indent


A classical result in black hole scattering is that in the low frequency limit, the absorption cross section of minimally coupled scalar fields by arbitrary spherically symmetric black holes is equal to the horizon area:
\be
\sigma_{\mathrm{cl}} = A_H. \label{seccaocl}
\ee
This has been shown first to four--dimensional spherically symmetric black holes \cite{u76}, and later generalized to $d-$dimensional black holes \cite{dgm96}, which have a metric of the form
\be \label{schwarz}
ds^2=- f(r) dt^2+ \frac{d r^2}{g(r)}+ r^2 d\Omega^2_{d-2}.
\ee
Also in \cite{dgm96} this result has been generalized to higher spin fields, but always in Einstein gravity.

In previous works \cite{Moura:2011rr,Moura:2006pz,Moura:2014epa} we have studied the scattering of $d-$dimensional nonextremal spherically symmetric black holes in the presence of leading string theoretical $\a$ corrections. Let $T_H$ is the black hole temperature\footnote{We designate the Hawking temperature by $T_H$ in order to more clearly distinguish it from other temperatures we will later consider.}, given from (\ref{schwarz}) by $T_H=\sqrt{\frac{g(R_H)}{f(R_H)}}\frac{f'(R_H)}{4 \pi},$ $R_H$ being the horizon radius, and let $\lambda$ represent the adequate power of $\a/R_H^2$ corresponding to those leading corrections. In \cite{Moura:2011rr} we have concluded that one always has, to first order in $\lambda,$
\be
T_H=\left.T_H\right|_{\a=0}  \left(1 + \lambda \ \delta T_H \right), \,
\sigma=\left.\sigma\right|_{\a=0} \left(1 - \lambda \ \delta T_H \right).
\label{deltat}
\ee
This means the leading $\a$ correction to such absorption cross section is symmetric to the one of the temperature. This $\a$ correction $\delta T_H$ is a number which depends on the concrete black hole solution one is considering, but also on the coordinate system. Indeed, (\ref{deltat}) was derived in a system of coordinates such that the horizon radius has no $\a$ corrections. In such system of coordinates, $\delta T_H$ in (\ref{deltat}) represents the only (explicit and implicit) $\a$ correction to the cross section. But the expression for the cross section should be covariant, which is certainly not the case of (\ref{deltat}). Although this equation can be used to express the cross section for every single particular solution, after having determined its $\a$-corrected temperature, it would certainly be desirable to have an expression for the $\a$-corrected cross section valid for every solution, like (\ref{seccaocl}) is for the classical cross section.

From (\ref{deltat}) we guess that, in order for the $(- \lambda \ \delta T_H)$ $\a$ correction to $\sigma$ to appear naturally and covariantly, to first order in the perturbative parameter $\lambda,$ it has to come implicitly from a $1/T_H$ factor in the respective formula, i.e., in order to verify (\ref{deltat}) the cross section must be of the form $\sigma = C/T_H,$ for some quantity $C$ depending on the solution and not having $\a$ corrections (for a formal proof see \cite{Moura:2011rr,Moura:2014epa}). This property of the quantity $C$ allows us to find it (and the cross section) just from classical solutions, even if the formula we are looking for is valid for $\a$-corrected solutions.

For a classical Tangherlini black hole the functions $f(r), g(r)$ in (\ref{schwarz}) are equal and given by $f(r)=g(r)=1 - \left(\frac{R_H}{r}\right)^{d-3}.$ The black hole horizon is a $(d-2)-$sphere with area $A_H = R_H^{d-2} \Omega_{d-2},$ $\Omega_{d-2}=\frac{2 \pi^{\frac{d-1}{2}}}{\Gamma\left(\frac{d-1}{2}\right)}.$ In this case, the horizon radius and the temperature are related by
\be
R_H(T_H)= \frac{d-3}{4 \pi T_H}. \label{rht}
\ee
Using this relation and taking the classical result (\ref{seccaocl}) for the cross section, but written in such a way that it has a $1/T_H$ explicit dependence, we are led to a covariant expression for the absorption cross section for Tangherlini--like (i.e. non--charged) black holes \cite{Moura:2011rr}:
\be
\sigma = \frac{d-3}{4 \pi T_H} \Omega_{d-2}^{\frac{1}{d-2}} A_H^{\frac{d-3}{d-2}}. \label{seccaoto}
\ee
Indeed, for a solution such that (\ref{rht}) is valid, one can easily check that (\ref{seccaocl}) is equi\-valent to (\ref{seccaoto}), if this last formula is considered classically (without its implicit $\a$ corrections). Although it does not have an explicit dependence on $\a,$ equation (\ref{seccaoto}) encompasses the $\a$ corrections to the classical result (\ref{seccaocl}): they come implicitly, through the $\a$ corrections to $T_H$ and $A_H$ (in a general system of coordinates, both corrections exist).

But in the presence of higher derivative terms the metric, and hence the horizon area, are subject to field redefinition ambiguities. As shown in \cite{Moura:2011rr}, it is impossible to express the horizon area, in a solution with $\a$ corrections, exclusively in terms of the (obviously also $\a$-corrected) mass or temperature in a way that is independent of the metric frame and of those field redefinitions: the corresponding expressions in terms of the mass or temperature in the string frame are different than those in the Einstein frame. The same is true for the $\a$-corrected absorption cross section (i.e. if one considers the $\a$ correction $(- \lambda \ \delta T_H)$ from (\ref{deltat})). In the same article, we also show that the only quantity which, classically, is proportional to the black hole horizon area and for which the expressions in terms of mass and temperature are the same in both frames is the Wald entropy. If the classical relation $\sigma= 4 G S$ (obtained directly from (\ref{seccaocl})) between the cross section and the Wald entropy was preserved by $\a$ corrections, the same would be true for the cross section. But we have shown that is not the case: $\a$ corrections to the entropy are different than those from the cross section.

This is not surprising: the minimally coupled scalar field obviously does not have a stringy origin; therefore, any physical property related to it, like the cross section, does not need to be invariant under metric frame transformations. On the contrary, the black hole metric is intrinsically stringy, and black hole properties such as the entropy should be invariant under frame transformations. Nonetheless, (\ref{seccaoto}) written in that form is valid for any metric frame.

Equation (\ref{seccaoto}) is therefore the only way to express the $\a$-corrected absorption cross section which is frame--independent, covariant and which includes naturally the $\a$ corrections - it is actually valid for every solution of the form (\ref{schwarz}). To summarize: the presence of $\a$ corrections introduces a new parameter in the black hole solution, precisely the inverse string tension $\a$ (or its power $\lambda$). As we have seen, keeping the dependence on $\lambda$ explicit does not allow for covariant and/or frame--independent expressions for the cross section. This way, one must find a way to make the dependence of the cross section on $\lambda$ just implicit and not explicit. Since $\lambda$ is a new parameter in the solution, an extra quantity must be present to provide the remaining $\lambda$ corrections to the cross section. Those corrections require such extra quantity (besides the horizon area) to be the black hole temperature, with an explicit dependence on it of the form of (\ref{seccaoto}), as we have seen.

This $1/T_H$ explicit dependence of (\ref{seccaoto}) is remarkable. As we mentioned, this formula has been derived for Tangherlini--like black holes, which do not have a finite extremal ($T_H \rightarrow 0$) limit. But, if a formula like (\ref{seccaoto}) is to be valid for more general black holes, allowing a valid finite extremal limit, as suggested in general in (\ref{deltat}), then the horizon area should have an implicit dependence on the temperature such that the absorption cross section remains finite in the limit $T_H \rightarrow 0$.

The two main questions we address in this article are therefore: Can a formula like (\ref{seccaoto}) for the cross section, with a $1/T_H$ explicit dependence, be obtained for more general (namely, charged) black holes? If so, is such formula well defined in the extremal limit?

The article is organized as follows. In section 2, we obtain the (simple) extension of formula (\ref{seccaoto}) to charged black holes. In section 3, we apply this extended formula to classical $d$-dimensional Reissner-Nordstr\"om black holes, just for illustrative purposes. We compare the obtained result for the low frequency cross section with the proposed formula of section 2, and we carefully analyze its extremal limit. In section 4 we discuss the application of formula (\ref{seccaoto}) to other classical stringy solutions, namely the D1-D5 system. In section 5 we apply our results to a spherically symmetric $d-$dimensional solution with $\a^3$ superstring corrections. We end by discussing our results.

\section{The absorption cross section for charged black holes (including string corrections)}
\label{secch}
\indent

In this section we give the main argument justifying the simple modification of (\ref{seccaoto}) generalizing it to charged black holes. The full details of the derivation can be found in \cite{Moura:2011rr,Moura:2014epa}.

We will keep dealing only with spherically symmetric black holes, but this time with some kind of charge. These black holes also have a metric of the form (\ref{schwarz}), but with
\be
f(r) = g(r) = c(r) \left(1 - \left(\frac{R_H}{r}\right)^{d-3}\right).
\label{tangherc} \ee
The black hole charges appear as parameters (which may or may not be independent) of the function $c(r).$\footnote{The most general $d-$dimensional spherically symmetric metric does not verify $f(r) = g(r)$: these appear as two independent functions. We take the slightly less general form (\ref{tangherc}) for simplicity, and because it is enough for the cases we consider in this article. In \cite{Moura:2014epa} we have studied the most general case.}
The temperature of these black holes is given by
\be
T_H=\frac{(d-3) c(R_H)}{4 \pi R_H}. \label{temph}
\ee
Clearly, extremal black holes are defined by the condition $c(R_H) \equiv 0.$

In order to find an expression equivalent to (\ref{rht}) for (\ref{tangherc}), first one must invert (\ref{temph}). That may or may not be possible, depending on the functional form of $c(R_H).$ But if we simply write (\ref{temph}) as
\be
R_H=\frac{(d-3) c(R_H)}{4 \pi T_H} \label{rhc}
\ee
and replace on (\ref{seccaocl}), we obtain
\be
\sigma = \frac{(d-3) c(R_H)}{4 \pi T_H} \Omega_{d-2}^{\frac{1}{d-2}} A_H^{\frac{d-3}{d-2}}, \label{seccaoc}
\ee
All the quantities ($T_H, A_H$ and also $c(R_H)$) in (\ref{seccaoc}) are expected to have $\a$ corrections; the corrections to the cross section result from the corrections (implicit in (\ref{seccaoc})) to these quantities (the derivation of (\ref{deltat}) in \cite{Moura:2011rr} allows for $c(R_H)$ being an $\a$-dependent constant). We see that in the formula for the cross section, for charged black holes, the necessary $1/T_H$ factor gets replaced by a $c(R_H)/T_H$ factor. In the extremal limit, both the numerator and the denominator of this factor go to 0, but from (\ref{temph}) we learn that $\frac{c(R_H)}{T_H} =\frac{4 \pi R_H}{d-3},$ a finite result.

For nonextremal black holes, as explained in \cite{Moura:2011rr}, one can always re-scale the time as $d\tilde{t}=c(R_H) \, dt$ (and equivalently, after euclideanization, the time periodicity, given by $1/T_H$). This means that, for nonextremal black holes, time and temperature can always be chosen in order to set $c(R_H) \equiv 1$, and therefore (\ref{seccaoc}) reduces to (\ref{seccaoto}). Counting the parameters, the discussion of the previous section for non-charged black holes also remains valid: for classical solutions, the cross section is given by one single quantity, the horizon area; including $\a$ corrections, one extra quantity, the black hole temperature, is necessary to express it, according to (\ref{seccaoto}).

Equation (\ref{seccaoc}) makes it evident that the low frequency absorption cross section is given as a quotient, where both the numerator and the denominator (the black hole temperature) vanish in the extremal limit, the whole expression remaining finite in such limit. If this expression is to be applied in such form (\ref{seccaoc}), one must verify that the functional form of $c(R_H)$ in terms of charges/mass is invariant under metric redefinitions: namely, that it looks the same in the string, Einstein and any other frame.

\section{The simplest example: the absorption cross section for the $d-$dimensional Reissner-Nordstr\"om black hole}
\label{rn}
\indent

We now apply our result from the previous section to the simplest charged black hole, the classical Reissner-Nordstr\"om solution. We recall that, for a classical solution like this, (\ref{seccaocl}) is perfectly applicable: (\ref{seccaoc}) only becomes necessary in the presence of $\a$ corrections. We just take this as an illustrative example on the study of the behavior at the extremal limit.

In $d$ dimensions, the classical Reissner-Nordstr\"om solution can be written in the form (\ref{schwarz}), with
\be
f(r)= g(r) = \left(1 - \left(\frac{R_Q}{r}\right)^{d-3}\right) \left(1 - \left(\frac{R_H}{r}\right)^{d-3}\right).
\label{rnd}
\ee
Without loss of generality, we take the inner horizon radius $R_Q$ not larger than the (event) horizon radius $R_H:$ $R_Q \leq R_H.$ The case $R_Q = R_H$ corresponds, as it is well known, to an extremal black hole, but in the following analysis we take nonextremal black holes. Clearly this solution is of the form (\ref{tangherc}). Its mass $M$ and electric charge $Q$ are given in terms of the radii $R_Q, R_H$ by
\bea
R_H^{d-3} = \mu + \sqrt{\mu^2 - q^2}, \qquad  R_Q^{d-3} = \mu - \sqrt{\mu^2 - q^2}, \nonumber \\
\mu = \frac{8 \pi }{\Omega_{d-2} (d-2)} M, \qquad q^2 = \frac{2}{(d-2)(d-3)}Q^2.  \label{mqrnd}
\eea

Its temperature is given by
\be
T_H=\frac{d-3}{4 \pi R_H} \left(1 - \left(\frac{R_Q}{R_H}\right)^{d-3}\right). \label{temprnd}
\ee

We see clearly what we previously mentioned for general charged black hole solutions -- it is not possible to express the horizon radius $R_H$ exclusively in terms of the temperature: by inverting (\ref{temprnd}), $R_Q$ always remains as an independent parameter, which was expressed in terms of $M$ and $Q$ in (\ref{mqrnd}).

The inversion of (\ref{temprnd}) is algebraically possible for $d=4, 5, 6$. For $d=4, 5$ we have
\bea
R_H^{d=4}(T_H)&=& \frac{1-\sqrt{1-16 \pi R_Q T_H}}{8 \pi T_H}, \nonumber \\
R_H^{d=5}(T_H)&=& \frac{1- \varrho_5(T_H) - \frac{1}{\varrho_5(T_H)}}{6 \pi T_H}, \nonumber \\
\varrho_5(T_H) &=& \left[54 \pi^2 R_Q^2 T_H^2+ 6 \sqrt{81 \pi^4 R_Q^4 T_H^4-3 \pi^2 R_Q^2 T_H^2}-1\right]^{\frac{1}{3}},
\eea
while the expression for $d=6$, although obtainable analytically, looks even more complicated and we prefer to omit it here. In any case, already from (\ref{temprnd}) we see that, at $T_H=0$, $R_H(T_H)=R_Q$. Differentiating (\ref{temprnd}) with respect to $R_H,$ from the inverse function theorem we can obtain $\frac{d\,R_H(T_H)}{d\,T_H}=\frac{1}{\frac{d\,T_H(R_H)}{d\,R_H}}.$ Taking this result at $T_H=0$ (or equivalently $R_H=R_Q$) allows us to write
\be
R_H(T_H)= R_Q + \frac{4 \pi R_Q^2}{(d-3)^2} T_H + {\mathcal{O}} \left(T_H^2\right).
\ee
This way we see that, for any $d$, $R_H(T_H)$ is regular at $T_H=0$, and it has a well defined extremal limit.

Concerning the invariance under field redefinitions, the transformation required to the change of metric frame results, for a spherically symmetric metric, in a change of scale in the radial coordinate: if the radial coordinates in two different frames (for example, but not strictly necessarily, the two most common Einstein and string frames) are $r_E, r_S,$ respectively, they are related simply by
\be
r_E = k\, r_S, \label{rers}
\ee
$k$ being an $\a$-dependent constant (for a detailed discussion see \cite{Moura:2011rr}). From (\ref{tangherc}) and (\ref{rnd}) we obtain $c(R_H)=1 - \left(\frac{R_Q}{R_H}\right)^{d-3},$ which only depends on the quotient $\frac{R_Q}{R_H}$. From what we have just seen, this quotient and therefore also $c(R_H)$ are frame--independent.

From the expression obtained above and (\ref{mqrnd}), we can then write $c(R_H)$ in terms of physical variables (mass and charge) as $c(R_H)=\frac{2 \sqrt{\mu^2 - q^2}}{\mu + \sqrt{\mu^2 - q^2}},$ although as discussed in the previous section one can always re-scale the time and temperature in order to have $c(R_H) \equiv 1.$ With such choice, the result for the low frequency absorption cross section of the $d$-dimensional Reissner--Nordstr\"om black hole is the same as the one for the noncharged black holes, given by (\ref{seccaoto}).

It is worth mentioning that formula (\ref{seccaoto}) is valid classically and up to the leading $\a$ corrections, even if such corrections to the Reissner--Nordstr\"om black hole, to our knowledge, have never been determined.

\section{The classical D1-D5 system}
\indent

Here we consider another classical solution, this time in the context of string theory (but still without string $\a$ corrections).

The D1-D5 system is a class of charged black hole solutions of type IIB supergravity compactified on $T^4 \otimes S^1.$  The D-brane description of these black holes involves a bound state of $Q_1$ fundamental strings  wrapping $S^1$ and $Q_5$ NS5-branes wrapping $T^4 \otimes S^1$. Excitations of this bound state are approximately described by transverse oscillations, within the NS5-brane, of a single effective string. These oscillations carry momentum and are described by a gas of left and right movers on the string, each with effective left and right moving temperatures $T_{L}, \, T_{R}.$ They are related to the overall Hawking temperature by $T_{H}^{-1} = \frac{1}{2}(T_{L}^{-1}+T_{R}^{-1})$.

In \cite{Das:1996wn} it was shown that the total rate of emission of low energy scalar quanta by this D1-D5 configuration is generically given by $$\Gamma(\omega)= g_{eff} \, \omega \, \rho\left(\frac{\omega}{2 T_L} \right) \, \rho \left(\frac{\omega}{2 T_R} \right) \frac{d^d k}{(2\pi)^d},$$
where $g_{eff}$ is a (charge-dependent but frequency-independent) effective coupling of left and right moving oscillations of energies $\omega/2$ to an outgoing scalar of energy $\omega$, and the thermal factor $\rho(\omega/T)$ is given by
\be
\rho \left(\frac{\omega}{T} \right)\equiv \frac{1}{e^{ \frac{\omega}{ T}}  -1}.
\ee
Together with the detailed balance condition
$$\Gamma(\omega)= \sigma_{abs}(\omega) \rho \left(\frac{\omega}{T_H} \right) \frac{d^d k}{(2\pi)^d}$$
we easily obtain
\be
\sigma_{abs}(\omega)=g_{eff} \, \omega \, \frac{\rho \left(\frac{\omega}{2 T_L} \right) \rho \left(\frac{\omega}{2 T_R} \right)}{\rho \left(\frac{\omega}{T_H} \right)} \simeq 4 g_{eff} \frac{T_L T_R}{T_H}  + {\mathcal{O}} \left( \omega \right). \label{seccaowg}
\ee
This way, we obtain for these solutions already in their classical, without $\a$ corrections limit, the same explicit dependence on the black hole temperature of the form $1/T_H$ that we got for $\a$-corrected black holes in (\ref{seccaoto}). The effective left and right moving temperatures $T_{L}, \, T_{R}$ should be expressed in terms of global charges. The numerator of the low frequency limit in (\ref{seccaowg}) must have an implicit dependence on $T_H$ in such a way that the cross section is finite in the extremal limit. Of course one could express that numerator explicitly in terms of $T_H$, but the resulting expression would be much more complicated than (\ref{seccaowg}), which was obtained naturally and has the adequate $T_H$ dependence to be valid including $\a$ corrections.

In \cite{Moura:2014epa} we considered different regimes of the D1-D5 system. For each case we computed the low frequency absorption cross section, showing the $1/T_H$ dependence and that the extremal limit of the cross section was well defined.

\section{Tangherlini-like black holes with $\a^3$ corrections in superstring theory}
\indent

In our previous works \cite{Moura:2011rr,Moura:2006pz} we have studied the scattering of $d-$dimensional spherically symmetric black holes with higher derivative corrections to first order in $\a$, from heterotic string theory. Now we apply it to a similar black hole, but with $\a^3$ corrections (quartic in the Riemann tensor) coming from type II superstring theory compactified on a flat $(10-d)$-torus. The corresponding effective action (in the Einstein frame) is given by
\bea
S &=& \frac{1}{16 \pi G} \int \sqrt{-g}\ \left( \R - \frac{4}{d-2} \left( \partial^\mu \phi \right) \partial_\mu \phi + \mbox{e}^{-\frac{12}{d-2}\phi} \frac{\zeta(3)}{16}\ \a^3\ Y (\R) \right) \, \mbox{d}^dx, \nonumber \\
Y (\R) &=& 2 \R_{\mu\nu\rho\sigma} \R^{\lambda\nu\rho\kappa} {\R^{\mu\alpha\beta}}_{\lambda} {\R_{\kappa\alpha\beta}}^{\sigma} + \R_{\mu\nu\rho\sigma} \R^{\lambda\kappa\rho\sigma} {\R^{\mu\alpha\beta}}_{\lambda} {\R_{\kappa\alpha\beta}}^{\nu}.
\eea
where $\zeta(s)$ is the Riemann zeta--function. All the gauge and fermionic fields can be consistently set to 0 - therefore there are no charges associated to this black hole, which is the Tangherlini solution with $\a^3$ corrections. The metric we consider was found in \cite{Myers:1987qx} and is of the form (\ref{schwarz}), with
\bea
f(r) &=& \left(1 - \left(\frac{R_H}{r}\right)^{d-3}\right) \left(1+ \frac{\zeta(3)}{8 R_H^6}\ \a^3 f_c(r) \right), \nonumber \\
g(r) &=& \left(1 - \left(\frac{R_H}{r}\right)^{d-3}\right) \left(1+ \frac{\zeta(3)}{8 R_H^6}\ \a^3 g_c(r) \right), \nonumber \\
f_c(r)&=& g_c(r) -C_d \left(\frac{R_H}{r}\right)^{3d-3}, \nonumber \\
g_c(r)&=& D_d \left(\frac{R_H}{r}\right)^{3d-3}+ E_d \frac{\left(\frac{R_H}{r}\right)^{2d}-1}{\left(\frac{R_H}{r}\right)^{2d} \left(\left(\frac{R_H}{r}\right)^{d-3}-1\right)}, \nonumber \\
C_d&=& \frac{32}{3}(d-3)(d-1) \left(2 d^3 - 10 d^2 +6d +15\right), \nonumber \\
D_d &=& \frac{2}{3} (d-3) \left( 52 d^4 -375 d^3 + 758 d^2 -117 d -570\right), \nonumber \\
E_d &=& -\frac{2}{3} (d-3) \left( 72 d^5 - 652 d^4 +2079 d^3 - 2654 d^2 + 837 d +570\right).
\label{myers}
\eea
The temperature of this black hole is given by
\be
T_H = \frac{d-3}{4 R_H \pi}\left[1- \frac{\zeta(3)}{16 R_H^6}\ \a^3 \left(C_d + 2 D_d + \frac{4d}{d-3} E_d \right)\right]. \label{temp}
\ee
This expression for the temperature is valid in the Einstein frame. In the string frame, the metric would still have the form (\ref{schwarz}), but the functions $f_c(r), g_c(r)$ would be different; we would have different values for the $\a^3$ corrections in (\ref{myers}) and, consequently, in $T_H$. Clearly the temperature, being a physical quantity intrinsic to the stringy black hole, would be the same, just with two different expressions in two different metric frames. Equating these expressions would give us the relation between the positions of the horizon radii in the different frames. That relation can be extrapolated to the radial coordinates, allowing us to obtain, for this solution, the constant $k$ in (\ref{rers}).

Inverting (\ref{temp}) one would obtain $R_H(T)$ which, replaced in (\ref{seccaoto}), gives an expression for the low frequency absorption valid only on the Einstein frame:
\be
\sigma = \Omega_{d-2} \left(\frac{d-3}{4 \pi T_H} \right)^{d-2} \left(1- \frac{\zeta(3)}{16}\ \a^3 \left(\frac{4 \pi T_H}{d-3} \right)^6 (d-2) \left(C_d + 2 D_d + \frac{4d}{d-3} E_d \right)\right).
\ee
On the other hand, as we mentioned, (\ref{seccaoto}) is valid in any metric frame.

\section{Conclusions}
\indent

In this article, we have shown that the low frequency absorption cross section for charged spherically symmetric $d$--dimensional black holes must have an explicit dependence on the black hole temperature of the form $1/T_H$. Such explicit dependence on a specific extra quantity (the temperature) appears as naturally induced (and required) by $\a$ corrections to the metric. It is manifest in formula (\ref{seccaoto}), valid for every spherically symmetric black hole including leading $\a$ corrections.

We took the classical $d$-dimensional Reissner-Nordstr\"om solution as a prototy\-pical example of the application of formula (\ref{seccaoto}). For this solution, we verified that, taking the cross section in the extremal limit results in a smooth behavior. But for extremal black holes nothing can depend on the temperature, and therefore formula (\ref{seccaoto}) is not valid and must be replaced. It remains an open issue to find a formula for the $\a$-corrected absorption cross section for extremal black holes. We address such issue in a forthcoming work \cite{Moura:2014lxa}.

Although we have derived formula (\ref{seccaoto}) assuming $\a$-corrected black holes (spherically symmetric, but not necessarily charged), we have also shown that, for a class of string-theoretical charged black holes (the D1-D5 system), such explicit dependence on the temperature of the form $1/T_H$ of the low frequency absorption cross section appears naturally already for $\a=0$, from different reasons and independently of our derivation.

We finally applied formula (\ref{seccaoto}) to a noncharged solution with $\a^3$ corrections from type II superstring theory.

\section*{Acknowledgments}
This work has been supported by FEDER funds through \emph{Programa Operacional Fatores de Competitividade -- COMPETE} and by Funda\c c\~ao para a Ci\^encia e a Tecnologia through projects Est-C/MAT/UI0013/2011 and CERN/FP/123609/2011.


\bibliographystyle{plain}

\end{document}